\documentclass[usenatbib]{mn2e}
\usepackage{amsmath,graphics,epsfig}
\usepackage{natbib,hyperref}
\title{Rotation, inflation, and lithium in the Pleiades}
\author[G. Somers \& M.H. Pinsonneault]
    {Garrett Somers and Marc H. Pinsonneault\\
    Department of Astronomy, The Ohio State University\\
    140 West 18th Ave, Columbus OH 43201, USA\\
    somers@astronomy.ohio-state.edu}

\newcommand\aj{{AJ}}%
\newcommand\araa{{ARA\&A}}%
\newcommand\apj{{ApJ}}%
\newcommand\apjl{{ApJ}}%
\newcommand\apjs{{ApJS}}%
\newcommand\aap{{A\&A}}%
\newcommand\aapr{{A\&A~Rev.}}%
\newcommand\aaps{{A\&AS}}%
\newcommand\mnras{{MNRAS}}%
\newcommand\pasp{{PASP}}%
\newcommand\ssr{{Space~Sci.~Rev.}}%
\newcommand\nat{{Nature}}%
\newcommand\gca{{Geochim.~Cosmochim.~Acta}}%
\begin{document}

\maketitle

\defcitealias{somers14}{SP14}

\def\teff   {{$T_{\rm eff}$}}
\def\teffs  {{$T_{\rm eff}s$}}
\def\msun   {{M$_{\odot}$}}
\def\wsun   {{$\omega_{\odot}$}}
\def\td     {{$\tau  _{D}$}}
\def\tds    {{$\tau  _{D}$s}}
\def\pi     {{$P _i$}}
\def\prot   {{$P_{\rm rot}$}}
\def\fk     {{$F_K$}}
\def\fc     {{$f_c$}}
\def\wc     {{$\omega_{crit}$}}
\def\ML     {{$\alpha_{\rm ML}$}}
\def\MLsol  {{$\alpha_{\odot}$}}
\def\delR   {{$\delta_{\rm R}$}}
\def\logg   {{$\log g$}}
\def\loggs  {{$\log g$s}}

\begin{abstract}
The rapidly rotating cool dwarfs of the Pleiades are rich in lithium relative to their slowly rotating counterparts. Motivated by observations of inflated radii in young, active stars, and by calculations showing that radius inflation inhibits pre-main sequence (pre-MS) Li destruction, we test whether this pattern could arise from a connection between stellar rotation rate and radius inflation on the pre-MS. We demonstrate that pre-MS radius inflation can efficiently suppress lithium destruction by rotationally induced mixing in evolutionary models, and that the net effect of inflation and rotational mixing is a pattern where rotation correlates with lithium abundance for $M_{*} < {\rm M}_{\odot}$, and anti-correlates with lithium abundance for $M_{*} > {\rm M}_{\odot}$, similar to the empirical trend in the Pleiades. Next, we adopt different prescriptions for the dependence of inflation on rotation, and compare their predictions to the Pleiades lithium/rotation pattern. We find that if a connection between rotation and radius inflation exists, then the important qualitative features of this pattern naturally and generically emerge in our models. This is the first consistent physical model to date that explains the Li--rotation correlation in the Pleiades. We discuss plausible mechanisms for inducing this correlation and suggest an observational test using granulation.
\end{abstract}

\section{Introduction} \label{sec:intro}

The classic Vogt--Russell theorem states that the mass and composition of a star uniquely determines its structure in hydrostatic equilibrium. Although this paradigm has proven to be a strong overall guide during the main sequence (MS), the implicit assumptions in the Vogt--Russell theorem break down in the pre-MS, where substantial star-to-star variations in luminosity are ubiquitous at fixed \teff\ in star-forming regions \citep[e.g.][]{hillenbrand97}. This behaviour has traditionally been interpreted as the signature of an intracluster age spread, but an alternative explanation is that stars at fixed mass and age can have non-uniform stellar parameters (i.e. radius) induced by processes neglected in traditional models, such as rotation, mass accretion, and magnetic fields. Much attention has been paid to understanding these processes, as a complete picture of pre-MS evolution is crucial for inferring the ages of young star-forming regions, measuring the low end of the stellar initial mass function, and constraining the distribution of circumstellar disc lifetimes, which in turn have important implications for both the angular momentum evolution of stars and the formation of planets. 

The light element $^7$Li is a sensitive diagnostic of pre-MS structure. Lithium burns through the $^7$Li$(p,\alpha)\alpha$ reaction at a temperature $T_{\rm Li} \sim 2.5 \times 10^6$~K, and is therefore directly destroyed in stellar envelopes when the temperature at the base of the surface convection zone ($T_{\rm BCZ}$) is high. During the period of deep convection on the pre-MS, $T_{\rm BCZ}$ surpasses $T_{\rm Li}$ for FGK stars, inducing Li depletion in the envelope \citep{iben65}. This `standard model' depletion terminates during the approach to the zero-age MS (ZAMS), when the convection zone retreats and again becomes cool. Because the rate of Li burning scales as $T^{20}$ near the destruction temperature \citep{bildsten97}, the magnitude of standard model depletion is extremely sensitive to $T_{\rm BCZ}$. In standard stellar models (SSMs), only mass and composition influence this temperature, so equal mass stars within clusters are expected to reach the ZAMS with equal surface Li abundances, and remain nearly fixed in A(Li)\footnote{A(Li) = 12 + log$_{10}$($n$(Li)/$n$(H))} until the MS turn-off.

By contrast, Li abundance spreads have been observed in many stellar populations, such as the late G and K dwarfs on the pre-MS and ZAMS \citep[e.g.][SP14 hereafter, and references therein]{somers14}, the mid-F dwarfs of 0.1--0.5 Gyr old clusters \citep[e.g.][]{boesgaard86,balachandran95}, and solar analogues of old clusters and the field (\citealt{sestito05} and references therein; \citealt{israelian09}). Various non-standard processes have been put forward to explain the development of these dispersions, but they are generally unable to simultaneously explain all three above classes. For example, rotational mixing can induce depletion and dispersion on the MS, but cannot replicate the pattern in young stars (see below). Explaining this multifaceted Li depletion pattern presents a serious challenge for stellar evolution theory, but its resolution must reveal interesting physics not included in the standard model.

\begin{figure}
\includegraphics[width=3.2in]{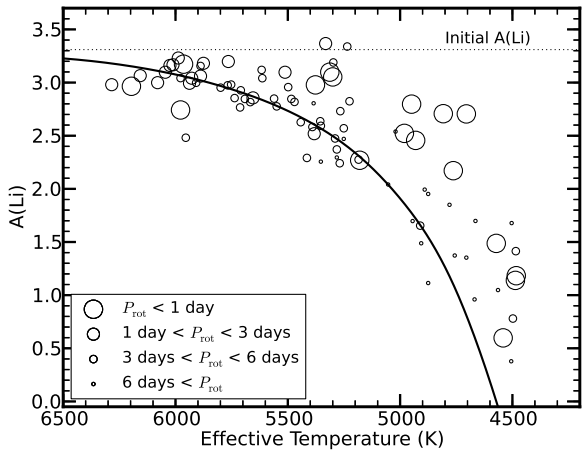}
\caption{The Li abundance pattern of the Pleiades (see $\S$\ref{sec:data}). The sizes of point represent stellar rotation periods, as indicated by the key, and clearly illustrate the enhanced Li abundance of rapid rotators redward of 5500K. The black line represents a standard stellar model prediction for the Li pattern of this cluster, taken from \citetalias{somers14}, and the dotted black line shows the initial Li abundance of these models.}
\label{fig:LiRot}
\end{figure}

In this work, we investigate a potential cause of the Li dispersion developing on the pre-MS. This dispersion has been identified in several young clusters, such as Blanco 1 and $\alpha$ Per \citep{jeffries99,balachandran11}, but we focus solely on the most notable example, the Pleiades. The Pleiades is a nearby open cluster \citep[d $\sim$ 133~pc;][]{an07}, just beyond the K dwarf ZAMS \citep[$t \sim 125$~Myr old;][]{stauffer98}, and with a near-solar iron abundance ([Fe/H] $\sim$ 0.03 $\pm$ 0.02; \citealt{soderblom09}). The Pleiades also exhibits a significant, \teff\ dependent spread in the $\lambda$6708 Li I absorption feature, originally identified by \citet{duncan83} and later confirmed by \citet{soderblom93}. This EW dispersion suggests an abundance spread at fixed \teff\ of a few tenths of a dex at 6000K, which increases to greater than an order of magnitude below 5500K (Fig. \ref{fig:LiRot}). This dispersion has remained unexplained since its discovery, but an intriguing clue has motivated several attempts at resolution: the most Li rich stars at fixed \teff\ tend to be the most rapidly rotating in the K dwarf regime \citep{soderblom93}. This observation has led several authors to suggest that the dispersion results from various observational biases related to rapid rotation, such as the enhancement of the $\lambda$6708 Li I feature due to blending from rotational broadening \citep{margheim02}, large spot filling factors and chromospheric activity altering observed equivalent widths \citep{soderblom93}, and non-LTE effects \citep{carlsson94}. However, \citet{king10} have shown that while these effects may contribute to the total dispersion, some underlying Li spread is likely present (see the discussion in $\S$6.2 of \citetalias{somers14}). It thus appears that some mechanism related to rotation affects Li depletion on the pre-MS.

The most widely discussed connection between rotation and Li is rotational mixing \citep[e.g.][]{pinsonneault90,pinsonneault92,zahn92,chaboyer95b,sestito05,li14}. Theory predicts that chemical materials can be exchanged between layers in the interiors of rotating stars due to several hydrodynamic processes, such as Eddington--Sweet circulation \citep{eddington29,sweet50,zahn92} and secular shear instabilities \citep[e.g.][]{zahn74}. Through these mechanisms, Li-depleted material from the deep interior can be mixed into the convective envelope, diluting the surface abundance and decreasing A(Li) below the standard model predictions. Rotational mixing has been invoked to explain the progressive destruction of Li observed in MS open clusters \citep[e.g.][]{pinsonneault97}, because it operates over time-scales analogous to MS lifetimes \citep{pinsonneault90}. It also naturally predicts the development of Li dispersions, found in some (but not all) old clusters \citep[e.g.][]{sestito05}, since depletion rates are determined by the non-uniform angular momentum (AM) histories of individual stars \citep[e.g.][]{irwin09}. Finally, rotational mixing decreases in efficiency at late ages due to stellar spin down, permitting the observed plateauing of Li abundances seen for old clusters \citep{sestito05}.

While rotational mixing may be responsible for MS depletion, it is unlikely to produce the pattern seen in the Pleiades K dwarfs. Li depletion through rotational mixing is most efficient in rapid rotators, opposite to the sense of the Li--rotation correlation in the Pleiades. This implies that some stronger process related to rotation on the pre-MS must counteract the depletion of Li through mixing. Several such mechanisms have been suggested in the literature, such as stronger internal shears in slow rotators due to disparate core--envelope coupling times \citep{bouvier08}, enhanced Li depletion in stars with extreme accretion histories \citep{baraffe10}, and enhanced internal shears due to long disc-locking times in slow rotators \citep{eggenberger12}. We note that each of these mechanisms induces a Li dispersion by enhancing the depletion of slow rotators, whereas the empirical calibration of \citetalias{somers14} favoured the suppression of depletion in rapid rotators.

\citetalias{somers14} proposed an explanation for the Pleiades Li spread: a dispersion in stellar radii at fixed mass and age during the pre-MS. If the convective envelope of a star were inflated relative to standard predictions, the temperature at the base of the convection zone would be reduced. Li destruction in the envelope is a strong function of this temperature, so stars of dissimilar radii would undergo different rates of Li depletion during the pre-MS, and thus arrive at the ZAMS with unequal Li abundances. This effect is so strong that a pre-MS radius spread of $\sim$10 per cent in non-rotating models produces a greater than order-of-magnitude spread in Li at the ZAMS \citepalias[][fig. 14]{somers14}, and if the most inflated stars were also the most rapidly rotating, the observed Li--rotation correlation would be naturally produced.

This idea was motivated by the rapidly-growing literature on discrepancies between the precisely measured radii of stars and the predictions of modern stellar evolution codes. Accurate radius measurements ($\sim$1--3 per cent) can be obtained by analysing the light curves of detached eclipsing binaries \citep[e.g.][]{popper97,torres10}, and by measuring the angular diameter of nearby stars with interferometry \citep[e.g.][]{boyajian12}. The average radii of open clusters can also be inferred by statistically modelling projected radius measurements ($R$ sin $i$; \citealt{jackson09,jackson14}). With each method, discrepancies between predicted and observed radii of $\sim$10 per cent have been found. These so-called `radius anomalies' have been identified, and found to correlate with rotation, in the pre-MS cluster IC 348 \citep{cottaar14}. They have also been claimed in young MS clusters, including the Pleiades \citep{jackson14}, and the magnitude of the discrepancy has been seen to correlate with chromospheric activity and rotation rate \citep{lopez-morales07,clausen09,stassun12,stassun14}.

In \citetalias{somers14}, we showed that a $\sim$10 per cent radius dispersion on the pre-MS can produce a Li dispersion commensurate with the Pleiades pattern. However, the impact of rotational mixing was not included. Mixing-driven Li destruction is strongest in rapid rotators, so this effect could compete with the inflationary inhibition of Li destruction. We therefore seek to assess whether a connection between rotation and radius inflation during the pre-MS represents a viable class of solutions to the Pleiades Li problem. To address this question, we simultaneously calculate the effects of rotational mixing and radius inflation on Li depletion in our evolutionary models, and compare the results to the empirical Pleiades distribution. To model radius inflation, we apply a modification to the mixing length in our stellar evolution code ($\S$\ref{sec:calib}). This approach provides us with a generic framework within which to explore the impact of radius on Li destruction. We construct a grid of models with dimensions of mass, radius inflation, and rotation using this method ($\S$\ref{sec:AMevol}), and collect a catalog of Pleiades data, for comparison with our models ($\S$\ref{sec:data}).

With this model grid, we explore the individual and simultaneous impacts of rotation and radius inflation on Li abundances ($\S$\ref{sec:section3}), and conclude that large inflation factors can indeed suppress rotationally-driven Li destruction in stellar models. Based on this result, we explore whether a putative connection between the two properties can produce a pattern which resembles the empirical Pleiades distribution ($\S$\ref{sec:section4}). Several limiting cases are ruled out, and a special case where faster rotation leads to greater radius inflation is found to compare favourably with the Pleiades. Thus, we conclude that the important qualitative features of the Pleiades Li and rotation pattern can be predicted in stellar models \textit{if and only if} radius inflation is correlated with rotation during the pre-MS. Detailed models tied to the underlying physical mechanism might well differ in their behavior; we discuss such models in $\S$\ref{sec:mechanism}. Finally, we suggest an observational test for the presence of a radius dispersion in the Pleiades and other clusters in $\S$\ref{sec:testing}, and summarise our conclusions in $\S$\ref{sec:conclusions}.

\section{Methods} \label{sec:methods}

The first step in our investigation is to generate a grid of stellar models with dimensions of mass, radius anomaly, and initial rotation conditions. Once complete, we can explore the generic behaviour of the models, and determine their ability to reproduce the Li--rotation correlation seen in the Pleiades. In this section, we first describe the stellar evolution code we employ, our basic calibration procedures, and our treatment of radius inflation. Next, we describe the implementation of rotation in our models. Finally, we assemble literature data for the Pleiades, as an empirical test of our models.

\subsection{Standard model calibration and radius inflation} \label{sec:calib}

Stellar models were calculated using the Yale Rotating Evolution Code (YREC; see \citealt{pinsonneault89} for a discussion of the mechanics of the code).  We adopt a present day solar heavy element abundance of $Z/X = 0.02292$ from \citet{gs98}, and choose the hydrogen mass fraction ($X$) and the mixing length coefficient (\ML) that reproduce the solar luminosity and radius for a 1\msun\ model at $4.568$~Gyr. The final solar calibrated values are $X = 0.71304$, $Z = 0.018035$, and \MLsol\ $= 1.88166$. Our models use the 2006 OPAL equation of state \citep{rogers96,rogers02}, atmospheric initial conditions from \citet{kurucz79}, high temperature opacities from the opacity project \citep{mendoza07}, low temperature opacities from \citet{ferguson05}, the $^7$Li$(p,\alpha)\alpha$ cross section of \citet{lamia12}, and the cross-sections of other nuclear reactions from \citet{adelberger11}. We treat electron screening using the method of \citet{salpeter54}, and do not include diffusion, as the timescales of this process are very long compared to the ages of systems we are concerned with. Our models are initialised with a Li abundance equal to the proto-solar abundance A(Li) = 3.31 \citep{anders89}.

We model radius inflation in this paper by reducing the mixing length parameter \ML, as suggested by \citet{chabrier07}. While inhibited convection can result directly from the presence of magnetic fields in the convective envelope of stars, we do not ascribe a specific underlying cause to the radius inflation considered in this paper. Due to the adiabatic nature of convective regions, a given fractional increase in the radius will have a similar effect on the temperature at the base of the envelope regardless of the underlying cause. We can therefore use an altered mixing length to explore the generic effects on Li abundances of any mechanism which increases the radii of stars. The mixing length is held fixed for each individual model during its lifetime, and is determined by requiring a specified inflation percentage at 10 Myr, an age characteristic of pre-MS Li destruction. This formulation embeds an age dependence of the radius anomaly in our models -- in particular, the anomaly remains nearly fixed as a function of age, even while stellar rotation rates evolve. This simplification may not capture important details of inflation/rotation connections tied to specific physical mechanisms, which require more sophisticated modeling to uncover. However, this framework is appropriate for investigating the general consequences of radius inflation on Li abundances.

Using the solar calibrated mixing length \MLsol, we calculated a baseline set of non-rotating models, with masses varying from 0.7\msun\ to 1.2\msun\ in steps of 0.05\msun, corresponding to the range of Pleiades members considered in this paper. These models are treated as possessing a radius anomaly of 0 per cent. We then define the function \delR($M$,\ML) as the fractional difference between the solar calibrated radius, $R(M,$\MLsol$)$, and the radius of another model of equal mass but mixing length \ML.

\begin{equation}
\delta_{\rm R}(M,\alpha_{\rm ML}) = \frac{R(M,\alpha_{\rm ML}) - R(M,\alpha_{\odot})}{R(M,\alpha_{\odot})}.
\label{eqn:delR}
\end{equation}

\begin{figure}
\includegraphics[width=3.3in]{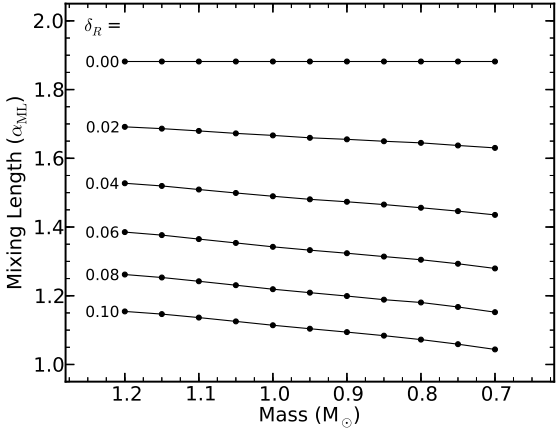}
\caption{The mixing length parameters \ML\ required to produce a radius anomaly \delR\ at 10 Myr, as a function of mass (see $\S$\ref{sec:calib}).}
\label{fig:infMLs}
\end{figure}

This value is calculated at 10 Myr as described above, an age characteristic of the epoch of pre-MS Li depletion in solar analogues. Next, we determined the \ML\ necessary to inflate a model of each mass to \delR($M$,\ML) = 0.02, 0.04, 0.06, 0.08, and 0.10, corresponding to radius anomalies of 2--10 per cent. The required \ML\ for each combination of mass and radius anomaly considered in this paper is shown in Fig. \ref{fig:infMLs}. A maximum anomaly of 10 per cent was chosen to coincide with the claimed radius anomalies in the Pleiades \citep{jackson14} and the upper envelope of anomalies observed in detached eclipsing binary systems and interferometric targets (see $\S$\ref{sec:intro}). 

We note that pre-MS Li depletion in stellar models is highly sensitive to the input physics. \citetalias{somers14} showed that adopting different published physical inputs, such as the equation of state or the $^7$Li$(p,\alpha)\alpha$ reaction cross-section, can produce significant changes in the expected pre-MS depletion of stars 0.9\msun\ and below. However, Li depletion on the MS is insensitive to physical inputs, so MS depletion predictions are robust once pre-MS depletion is correctly predicted. To address this issue, \citetalias{somers14} performed an empirical calibration of their models using the warm (\teff\ $>$ 5500K) stars in the Pleiades (see \citetalias{somers14} $\S$3--4 for details), and derived systematic corrections to the ZAMS Li abundance as a function of \teff. The corrections range from $\sim$0.2 dex in A(Li) at 6000K to $\sim$1 dex at 5000K. In this work, we apply these `empirically calibrated' pre-MS Li depletion corrections to our models. 

\subsection{Rotation and angular momentum evolution} \label{sec:AMevol}

To initialise rotation, we first evolve non-rotating models down the Hayashi track to the deuterium birth line, defined as the age at which 1 per cent of the initial D abundance has been destroyed ($\sim$ 0.1~Myr for the Sun). At this age, the launch conditions are applied. We assign an initial rotation period \pi, and assume this period to be fixed for a specified time \td, though magnetic interactions with a circumstellar disc \citep[e.g.][]{koenigl91}. Longer disc-locking times lead to slower subsequent rotation rates, as the stars are prevented from spinning up until a later epoch. Thereafter the rotation rate evolves under the influence of AM loss at the surface through magnetised winds. To model the loss of angular momentum, we adopt a modification of the wind law of \citet{kawal88}, first proposed by \citet{krishnamurthi97}. In this formulation, the loss rate scales with the cube of the angular velocity, up to a critical `saturation' velocity \wc, when the loss law becomes linear with $\omega$. The saturation threshold \wc\ is scaled by the Rossby number (rotation frequency $\omega$ times the convective overturn time-scale $\tau_{CZ}$) of each model, which is normalised on the Sun ($\omega_{\odot} \tau_{CZ,\odot} \sim 2.2$). The formula is given by Eq. \ref{eqn:kawaler},

\begin{equation} 
 \displaystyle \frac{dJ}{dt} = \left\{
	\begin{array}{l l}
	 \displaystyle F_K \left(\frac{R}{R_{\odot}}\right)^{\frac{1}{2}} \left(\frac{M}{M_{\odot}}\right)^{-\frac{1}{2}} \omega \left( \frac{\omega_{crit}}{\omega_{\odot}} \right)^2, \quad \omega_{crit} < \omega \frac{\tau_{CZ}}{\tau_{CZ,\odot}} \\
	 \displaystyle F_K \left(\frac{R}{R_{\odot}}\right)^{\frac{1}{2}} \left(\frac{M}{M_{\odot}}\right)^{-\frac{1}{2}} \omega \left( \frac{\omega \tau_{CZ}}{\omega_{\odot} \tau_{CZ,\odot}} \right)^2, \quad \omega_{crit} \geq \omega \frac{\tau_{CZ}}{\tau_{CZ,\odot}} \\
	\end{array} \right. 
	\label{eqn:kawaler}
\end{equation}

\fk\ represents the normalisation of the magnetic field strength, which is typically calibrated to reproduce observational data. Convection zones are assumed to rotate as a solid body, whereas radiative zones conserved angular momentum locally. The redistribution of angular momentum and chemical species throughout the radiative zones of star occurs purely through hydrodynamic instabilities in these models \citep[e.g.][]{endal78}, and are calculated by solving the coupled differential equations,

\begin{equation} 
\rho r^2 \frac{I}{M} \frac{d \omega}{dt} = f_{\omega} \frac{d}{dr} \left( \rho r^2 \frac{I}{M} D \frac{d\omega}{dr} \right),
\label{eqn:amt1}
\end{equation}

\begin{equation}
\rho r^2 \frac{d X_i}{dt} = f_c f_{\omega} \frac{d}{dr} \left( \rho r^2 D \frac{dX_i}{dr} \right).
\label{eqn:amt2}
\end{equation}

Here, I/M is the moment of inertia per unit mass, D is the characteristic diffusion coefficient of the angular momentum redistribution processes \citep{pinsonneault89,pinsonneault91,chaboyer95a}, $f_\omega$ is a scaling factor that sets the efficiency of this transport ($f_\omega$ = 1 in our models), and \fc\ is a scaling factor that sets the efficiency of transport of chemical species $X_i$ relative to $f_\omega$ \citep{chaboyer92}. 

Our rotation and mixing formulation leaves us with three free parameters: \fk, $\omega_{\rm crit,\odot}$, and \fc. To calibrate the two rotation parameters, \fk\ and $\omega_{\rm crit,\odot}$, we adopt the standard technique of fitting the median and rapid rotating envelopes of empirical distributions \citep[e.g.][]{vansaders13}. As Li depletion begins as early as a few Myr for higher mass stars, it is necessary to initialise our models near the birth-line. However, clusters around this age are unreliable rotation distributions benchmarks as many of their members retain their primordial circumstellar discs, which can exchange angular momentum with their host stars in complicated ways. Instead we select an older cluster, whose circumstellar discs have dissipated, and solve for the range of initial conditions necessary to evolve towards its rotation distribution.

For this purpose, we choose the 13 Myr old cluster h Per \citep{moraux13}. This cluster provides a useful benchmark for generating initial conditions as it lies on the pre-MS, and is old enough that gaseous accretion discs are no longer common. We fix the initial rotation period \pi\ = 8 days for all models, and find that \td\ = 3 Myr accurately reproduces the median rotation period in h Per, and that \td\ = 0.1 Myr accurately reproduces the 90th percentile of rapid rotators, regardless of mass. While an equal initial rotation period for all models may appear to be a crude approximation for an initial rotation distribution, there exist strong degeneracies in the choice of \pi\ and \td; in the absense of radius inflation, a fast spinning star with a long disk-locking time will be indistinguishable in both rotation and Li from a slowly spinning star with a short disk-locking time, once de-coupled from their disks \citep[e.g.][]{denissenkov10}. Even if differential radius anomalies (from 0-10 per cent) correlate with rotation before the age of h Per, we estimate that the resulting Li abundance dispersion will be $\sim$0.2 dex at 13 Myr in the most extreme case, and geneally much less. This dispersion is comparable to systemtics in the abundance determination, and far smaller than the Li dispersion our models seek to explain ($\S$\ref{sec:section4}). We choose \td\ = 6 Myr as the upper limit for disc-locking times, in accordance with the observed upper limit of disc survival lifetimes \citep{haisch01,fedele10}.

With our launch conditions set, we calibrated \fk\ and $\omega_{\rm crit,\odot}$ on the 550 Myr old open cluster M37. We obtained the M37 rotation catalog from \citet{hartman09}, and determined both the median rotation rate of the full sample, and the median rotation rate of the most rapidly rotating 20 per cent, as a function of \teff. We then evolved stellar models using median and rapid launch conditions (\td\ = 3 and 0.1 Myr, respectively), with a variety of values for \fk\ and $\omega_{\rm crit,\odot}$, to the age of the M37. The combination of parameters that best fit the cluster rotation pattern are \fk\ = 3.3 and $\omega_{\rm crit,\odot}$ = 9.5\wsun. These parameters are broadly consistent with values obtained by other authors \citep[e.g.][]{krishnamurthi97}. With the rotation parameters set, \fc\ was determined by requiring a median rotating solar mass model to match the median, metallicity-unbiased Li measurements of the Hyades from \citetalias{somers14}, a cluster of similar age to M37. With this method, we obtain \fc\ = 0.05. This value is consistent with typical results for rotational mixing calculations in the literature \citep[e.g.][]{pinsonneault90} and with anisotropic turbulence expected from theory \citep{chaboyer92}.

With the necessary values of \ML\ in hand, and the angular momentum and chemical transport calibrated, we calculated a grid of \pi\ = 8 day, solar metallicity models, with dimensions of mass (0.7--1.2\msun), \delR\ (0.0--0.1), and \td\ (0.1--6 Myr).

\subsection{Pleiades data} \label{sec:data}

To construct a comparison data set of \teff, A(Li), and \prot\ for the Pleiades, we draw \teffs\ and A(Li)s from \citetalias{somers14}, and cross-correlate this sample with the Pleiades rotation catalog of \citet{hartman10}. The resulting data can be seen in Fig. \ref{fig:LiRot}. We note that the Pleiades K dwarfs are known to show colour abnormalities possibly related to surface magnetic activity, radius inflation, and star spots \citep{jones72,vanleeuwen87,stauffer03}. These effects may also impact the measurement of the $\lambda$6707.8 Li I equivalent width, and thus the inferred abundance. However, these effects have little impact on the forthcoming results, as we seek to understand only the generic features of the empirical pattern, which appear secure (see $\S$\ref{sec:intro}), and not the detailed temperatures or abundances of the individual stars.

\section{The impact of rotation and inflation} \label{sec:section3}

Before performing a direct comparison between the models described in $\S$\ref{sec:methods} and the empirical Pleiades pattern, we explore the generic impact of inflation and rotation on the destruction of Li during the pre-MS in the models. We begin by reviewing the individual consequences of these two effects on stellar Li abundances in $\S$\ref{sec:roteffects}--\ref{sec:infeffects}. Rapid rotation is shown to \textit{enhance} the rate of Li depletion due to the strong influence of rotational mixing, and radius anomalies are shown to \textit{inhibit} the rate of Li depletion due to a decreased temperature at the base of the convection zone. Next, we investigate models with both rotation and inflation in $\S$\ref{sec:combeffects}. The resulting pattern is more complicated: Li depletion can either increase or decrease compared to a standard benchmark (no inflation, no rotation) depending on the relative strength of the two effects. In particular, we find that if a connection between rotation and inflation is present, Li abundance tends to correlate with rotation in cool stars. This motivates an exploration of the observational signature of different plausible inflation/rotation relationships in the next section. Readers interested only in the final results may proceed to $\S$\ref{sec:section4}.

\begin{figure*}
\includegraphics[width=4.5in]{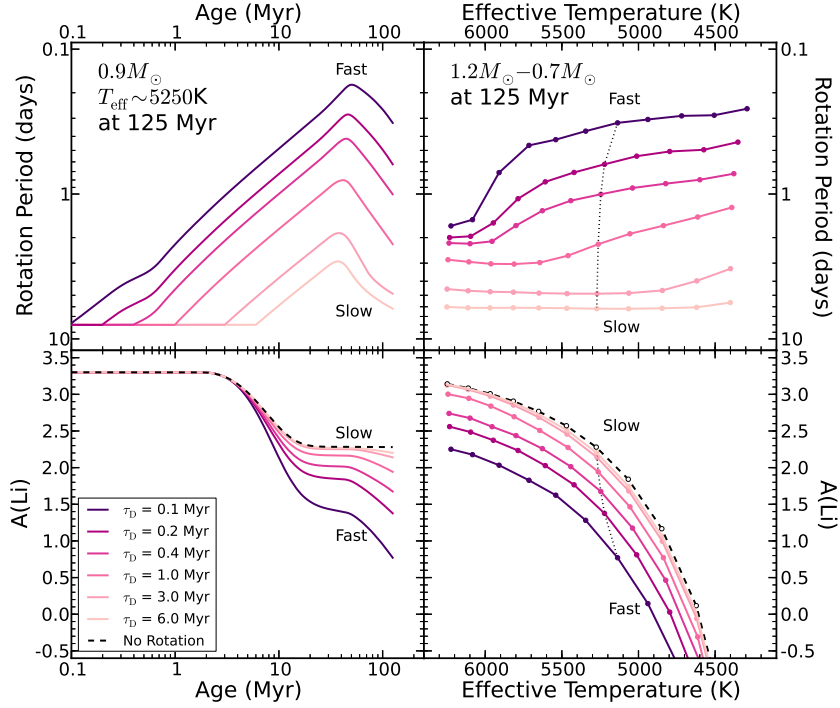}
\caption{The impact of rotation on Li depletion in non-inflated models. \textit{Left:} The evolution of the surface rotation period (top) and the envelope Li abundance (bottom) in 0.9\msun\ non-inflated models. A range of initial disc-locking times produces a range of Li depletion factors at the age of the Pleiades, which mainly set in between $\sim$5--20 Myr. Faster rotation leads to greater depletion in these models. \textit{Right:} Isochrones of rotation and Li at the age of the Pleiades. A large range of rotation rates persist at this age, given our initial rotation distribution. A Li dispersion forms at all temperatures in the models, and fans out below the standard prediction (black dashed line). A black dotted line shows the location of the 0.9\msun\ models represented in the left column.}
\label{fig:InflTracksRot}
\end{figure*}

\subsection{Rotation} \label{sec:roteffects}

In the theory of rotational mixing, meridional circulation and internal shears transport material between layers, leading to a gradual dilution of the convective envelope with Li-depleted material from the deep interior. Stars are born with a range of initial rotation conditions, so the rate of mixing differs from star to star in young systems, leading to the emergence of Li dispersions. Both the convective envelope and the radiative interior spin down over the course of 0.5--5 Gyr, eventually losing memory of their initial conditions, and asymptoting to rotation profile that depends on age but not early rotation rate \citep[e.g.][]{krishnamurthi97}. This process equalises the rate of late-time Li depletion between stars with different AM histories, thus freezing out the intra-cluster Li dispersion. Throughout stellar lifetimes, the strength of mixing depends both on the absolute rotation rate of the star and on the degree of differential rotation within the star; this makes Li depletion efficient at young ages and inefficient at old ages, given the progressive spin-down of stars. This generic picture of MS rotational mixing has been long established, but we are concerned in this work with the effect of rotation on the pre-MS, which is often neglected in mixing studies due to its small assumed effect relative to standard model depletion. We therefore present an exploration of the impact on rotation on Li during the pre-MS.

The top left of Fig. \ref{fig:InflTracksRot} shows rotation tracks for several solar calibrated (\delR\ = 0.0), 0.9\msun\ models with different disc-locking times. The models remain at constant rotation period (8 days) until detaching from their discs, when pre-MS contraction begins to increase their rotation rate. Once the models fully contract on to the ZAMS ($\sim$ 50~Myr), they begin to spin down under the influence of magnetic winds. Shorter disc-locking times lead to faster rotation at the ZAMS, given an equal initial rotation rate. Thus the observational result that disc lifetimes differ from star to star naturally implies a $>$1~dex range of rotation periods at the ZAMS. The bottom left panel follows the evolution of Li in the envelope of each model. Three stages of pre-MS Li evolution are revealed: 1) no depletion for the first few Myr; 2) an epoch of depletion from $\sim$ 3--20 Myr; 3) very little depletion on the late pre-MS. Thereafter, rotational mixing reduces A(Li) on the MS. The second pre-MS phase represents the time when the temperature at the base of the convection zone surpasses the Li depletion temperature (see $\S$\ref{sec:intro}). The rate of Li destruction during this phase is a strong function of rotation, so depletion in rapid rotators is high. This is because rotational mixing is particularly efficient when the material directly below the convection zone is highly depleted, as is the case when convection zones are deep. We have also plotted the Li track of a standard, non-rotating model (black dashed line) for comparison. This model depletes nearly the same amount of Li during the pre-MS as the slow rotating models (\td\ $\ga$ 1 Myr), suggesting that very rapid rotation ($\la$ 1 day period) is necessary for an appreciable enhancement to the rate of pre-MS depletion. 

\begin{figure*}
\includegraphics[width=4.5in]{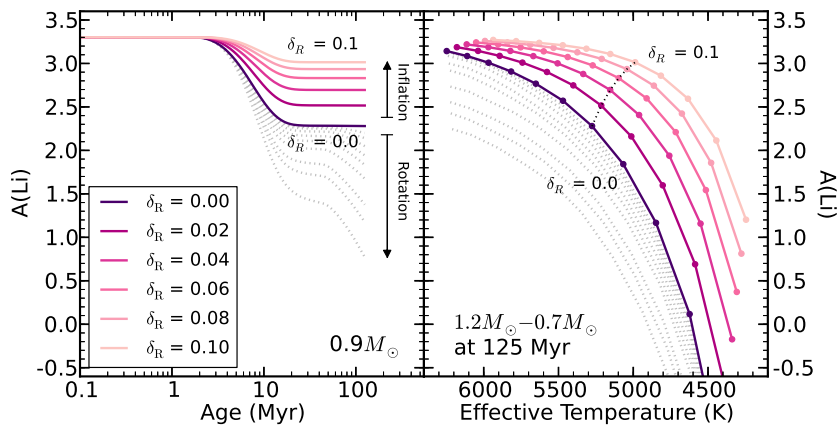}
\caption{The impact of inflation factors of 2-10 per cent on Li depletion in non-rotating models. \textit{Left:} Li tracks of 0.9\msun\ models. A greater radius anomaly decreases the rate of depletion during the 5--20 Myr age range, imparting a dispersion at 125 Myr. This dispersion lies above the standard prediction, opposite of the rotationally-induced dispersion (grey dotted lines). \textit{Right:} Inflated isochrones at the age of the Pleiades.  A large dispersion is present which increases towards lower temperatures. The black line in this figure is identical to the black dashed line in Fig. \ref{fig:InflTracksRot}.}
\label{fig:InflTracksInf}
\end{figure*}

The right column of Fig. \ref{fig:InflTracksRot} explores the mass dependence of this behaviour at the age of the Pleiades. In the top panel, each individual line represents a fixed disc-locking time, and the points signify mass steps of 0.05\msun, with the 0.9\msun\ locus indicated by a dotted line. A large range of rotation rates persist at 125 Myr within our mass range. The rotation dispersion is largest below $\sim$ 5700K, and decreases towards larger \teff. We also see that the structural effects of rapid rotation lead to a reduction of the surface temperature by $\sim$ 50--150K, independent of spot-induced changes. This results from the additional pressure support provided by rotation at the equator, which leads to a lower effective gravity, and hence a larger equilibrium radius. The bottom right panel shows Li isochrones at the age of the Pleiades. Lower mass stars undergo deeper convection during the pre-MS, and spend more time on the pre-MS, resulting in more rapid Li destruction over an extended period of time, and ultimately a lower abundance at 125 Myr. Variable rotation induces a Li dispersion commensurate with the 0.9\msun\ spread at all masses in the considered range, suggesting that the 0.9\msun\ Li tracks are representative of the behaviour of Li destruction in all of our models. Thus, our rotating models predict that at the age of the Pleiades, slow rotators should cluster near the standard prediction (dashed black line), and fast rotators should be depleted below the standard prediction by $\sim$1.5 dex.

These results are necessarily dependent on the details of the stellar models, notably the treatment of angular momentum transport and loss. While we have assumed that angular momentum is redistributed purely by hydrodynamics and mechanics ($\S$\ref{sec:intro}), some authors have argued that empirical rotation trends are better fit by models which include additional transport processes operating in the interior (e.g. \citealt{bouvier08}; \citealt{denissenkov10}; \citealt{gallet13}), which could have implications for pre-MS Li destruction. A full treatment of this additional physics is outside the scope of this project, but we note that since rotational mixing can only destroy Li, and not inhibit its destruction, the non-rotating case presented in Fig. \ref{fig:InflTracksRot} represents an approximate upper limit to the abundance at a given age (it is not an exact upper limit, as the minor structural effects of rotation described above can cool the base of the convection zone somewhat; e.g. \citealt{eggenberger12}).

\subsection{Inflation} \label{sec:infeffects}

\begin{figure*}
\includegraphics[width=4.5in]{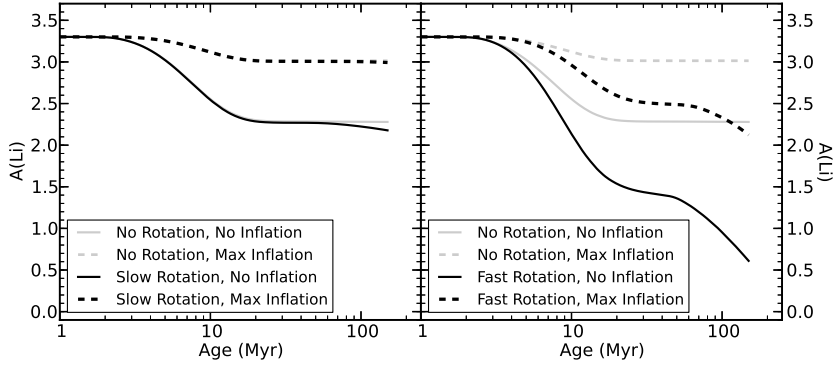}
\caption{\textit{Left:} The simultaneous effects of inflation and slow rotation. A standard model (solid grey), an inflated non-rotating model (dashed grey) and a non-inflated, slowly rotating model (solid black) are compared to an inflated, slowly rotating model (dashed black). Inflation dominates the evolution of Li when rotation is negligible. \textit{Right:} Same as left panel, except with rapid rotation instead of slow rotation. The depletion effects of maximal inflation (10 per cent) and maximal rotation (\td = 0.1 Myr) nearly cancel each other out at the age of the Pleiades, as the inflated rapid rotator (dashed black) reaches a similar abundance as the standard model (solid grey) at 125 Myr. It is clear that both inflation and rotation can strongly influence Li depletion. }
\label{fig:Li_tracks}
\end{figure*}

\begin{figure*}
\includegraphics[width=7.0in]{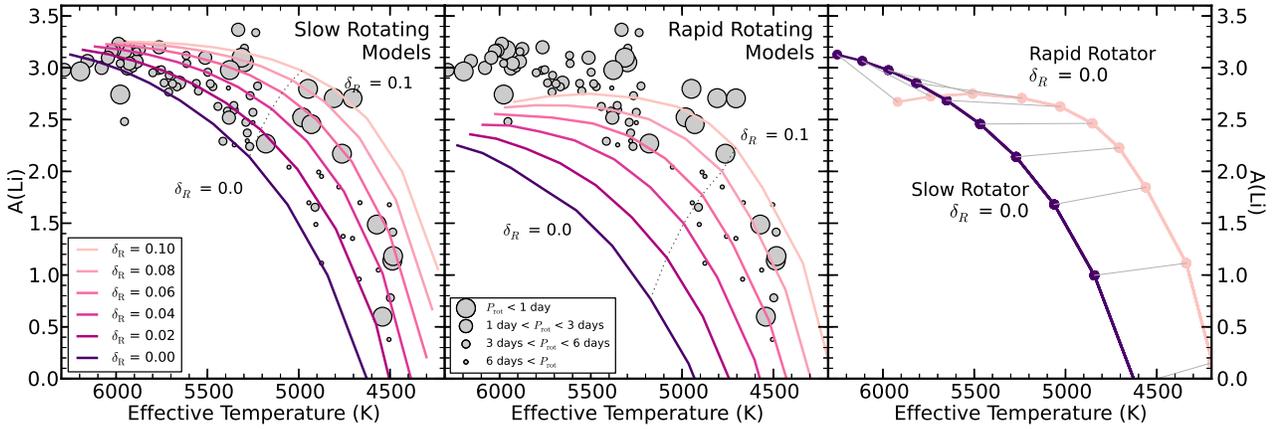}
\caption{Pleiades-age isochrones for models with both inflation and rotation. \textit{Left:} Slow rotating models with a variety of inflation factors. The Pleiades data from Fig. \ref{fig:LiRot} are also plotted. The Li dispersion resulting from a 10 per cent anomaly in slow rotators matches the observed dispersion as a function of \teff\ well. \textit{Centre:} Same as left panel, but with rapid rotation. The most Li-rich, rapidly rotating Pleiads coincide with the maximally inflated, rapidly rotating isochrones. \textit{Right:} The slow rotating, non-inflated isochrone from the left panel and the rapidly rotating, maximally inflated isochrone from the central panel are reproduced, along with lines matching models of equal mass. Below 5500K, these models succeed at producing the inverted Li--rotation pattern, as seen in the Pleiades.}
\label{fig:Li_isochrones}
\end{figure*}

Next, we assess the impact of radius inflation in the absence of rotational mixing. The left panel of Fig. \ref{fig:InflTracksInf} shows in colour the Li tracks of non-rotating 0.9\msun\ models with a variety of mixing lengths, which induce the radius anomalies listed in the key. The three stages of pre-MS depletion described above persist in inflated models, but the magnitude of depletion in the second stage is found to be highly sensitive to the stellar radius. Inflation reduces the temperature at the base of the convection zone, suppressing the rate of Li proton capture reactions and leading to a higher ZAMS abundance. The non-inflated, rotating tracks from Fig. \ref{fig:InflTracksRot} are shown in dotted grey for comparison. While increasing rotation \textit{enhances} the destruction of Li, increasing inflation \textit{inhibits} the destruction of Li. Both effects induce a dispersion, but the rotationally-induced spread lies below the standard prediction, and the inflationary spread lies above the standard prediction. The inflated models show no signs of additional depletion once on the MS, due to the lack of rotational mixing.

The right panel shows inflated isochrones at the age of the Pleiades, with the 0.9\msun\ locus denoted by the black dotted line. Li dispersions emerge generically at all masses, as with rotation, but the magnitude of the dispersion imparted by inflation is a stronger function of temperature: the dispersion increases from $\sim$ 0.92 dex at 6000K to $\sim$ 1.30 dex at 5000K due to rotation, while increasing from $\sim$ 0.26 dex to 1.40 dex at the same temperatures due to inflation. Importantly, the dispersion at fixed \teff\ is larger than the dispersion at fixed mass; radius inflation substantially reduces the \teff\ of stars, so the \teff\ of inflated, higher mass, Li-rich stars can be equal to that of non-inflated, lower mass, Li-poor stars. Ultimately, we see that a spread in radii of 10 per cent during the pre-MS can produce a Li dispersion that is small at temperatures typical of early G stars ($\sim$ 6000K), and increases to nearly two orders of magnitude at temperatures typical of mid K dwarfs ($\sim$ 4500K). These results are qualitatively similar to the findings of \citetalias{somers14}, who examined isochrones of constant mixing length instead of constant radius anomaly. 

To summarise $\S$\ref{sec:roteffects}--\ref{sec:infeffects}, increased rotation and increased inflation have the opposite effect for Li depletion: faster rotation leads to lower Li abundances, while higher inflation leads to greater Li abundances. Both effects can induce dispersions if they vary from star to star, but inflationary dispersions lie above standard predictions, and rotational dispersions lie below standard predictions. However, both effects serve to reduce stellar \teffs\ below standard stellar model predictions, due to the increasing radius driven by rotational pressure support and reduced convective efficiency.

\subsection{Rotation and inflation} \label{sec:combeffects}

We now present models which include both rotation and inflation. In this section, we permit any combination of these two parameters, for the purposes of exploring their combined impact on Li. Fig. \ref{fig:Li_tracks} compares the tracks of rotating models with a 10 per cent radius anomaly (dashed black; slowly rotating in the left panel, rapidly rotating in the right) to three models from $\S$\ref{sec:roteffects}--\ref{sec:infeffects}: a non-inflated, non-rotating model (solid grey), a model with a 10 per cent radius anomaly and no rotation (dashed grey), and a rapidly rotating model with no inflation (solid black; slowly rotating in left, rapidly rotating in right). 

In the left panel, the inflated slow rotator track (dashed black) is nearly identical to the inflated non-rotator track (dashed grey), suggesting that the inflationary suppression of Li destruction dominates when rotation is mild. By contrast, the right panel shows that rapid rotation induces \textit{significant} extra depletion relative to the non-rotating case in the presence of inflation. The non-rotating, inflated model (dashed grey) is significantly more abundant than the rapidly rotating, inflated model (dashed black), which in turn is far more abundant than the rapidly rotating, non-inflated model (solid black). Interestingly, the final abundance of the rapidly rotating, inflated model (dashed black in right panel) is close to that of the slowly rotating, non-inflated model (solid black in the left panel). This demonstrates that the counteracting effects of rotational mixing and inflation can almost entirely cancel each other out with regard to Li destruction, allowing a rapidly rotating inflated star to possess the same Li abundance as a slowly rotating non-inflated star at the age of the Pleiades. 

To explore the mass dependence of this behaviour, we present isochrones of slow and rapid rotation at 125 Myr in the left and central panels of Fig. \ref{fig:Li_isochrones}. Pleiades data are also plotted as circles, and the dashed line signifies the locus of 0.9\msun\ models. The slow rotating models make similar predictions to the inflated, non-rotating models presented in Fig. \ref{fig:InflTracksInf}, again demonstrating that the structural and hydrodynamic effects of rotation are minimal at low angular velocity. The abundance spread resulting from a 10 per cent radius anomaly matches the Pleiades dispersion well, as did our non-rotating models in \citetalias{somers14}. Compared to these models, the rapid rotating isochrones in the central panel are significantly depleted. The 0 per cent anomaly isochrone is so heavily depleted by rotational mixing that it lies far below the data. However, the inflation of the 10 per cent anomaly isochrone suppresses rotational mixing enough to approximately trace the locus of rapid rotators below $\sim$ 5000K. This correspondence is driven in part by the suppression of Li depletion, and in part by the substantial reduction in the surface temperature of our models, demonstrated by the dotted line (0.9\msun\ locus). 

Finally, the right panel of Fig. \ref{fig:Li_isochrones} reproduces the slow rotating, non-inflated isochrone from the left panel in purple, and the rapidly rotating, inflated isochrone from the central panel in peach. The result is intriguing: below $\sim$ 5700K, the rapid rotators appear \textit{more abundant} in Li than the slow rotators. This is the opposite sense from the pure rotation case, where faster rotation always leads to greater Li depletion, but the same sense as the pattern observed in the Pleiades. Furthermore, the dispersion is small around 5500K, but increases substantially towards lower temperatures until reaching nearly 2 dex at 4500K, also reminiscent of the empirical Pleiades pattern. This behaviour suggests that a radius inflation which increases with rotation may be able to both produce the correct sense of the Li--rotation correlation seen in the Pleiades, and accurately predict the width of the dispersion as a function of temperature. In order to assess this possibility, we consider in the following section the observational patterns resulting from different functional dependencies of inflation on rotation. 

\section{Sleuthing the rotation--inflation connection} \label{sec:section4}

In $\S$\ref{sec:combeffects}, we argued that certain qualitative features of the Pleiades Li pattern are predicted if rapidly rotating stars are inflated relative to their slow rotating counterparts. In this section, we expand upon this idea by generating synthetic cluster distributions with a variety of inflation/rotation prescriptions, initialised to reproduce the Pleiades rotation distribution, and comparing them directly to the data.

To construct these synthetic distributions, we first adopt a rule that explicitly links the degree of inflation to the rotation period at a benchmark age of 10 Myr [\delR\ = \delR(\prot)]. Using this rule, we collapse our 3-dimensional model grid (mass, \td, inflation) into two dimensions (mass, \td) by interpolating between models of fixed $M$ and \td, but variable \delR, to the location in parameter space where the inflation rule \delR\ = \delR(\prot) is satisfied. The result is a specific mapping between each choice of $M$ and \td\ (corresponding to a specific \prot\ at 10 Myr), and the relevant observables, \teff, A(Li), and \prot, at 125 Myr.

\begin{figure*}
\includegraphics[width=7.0in]{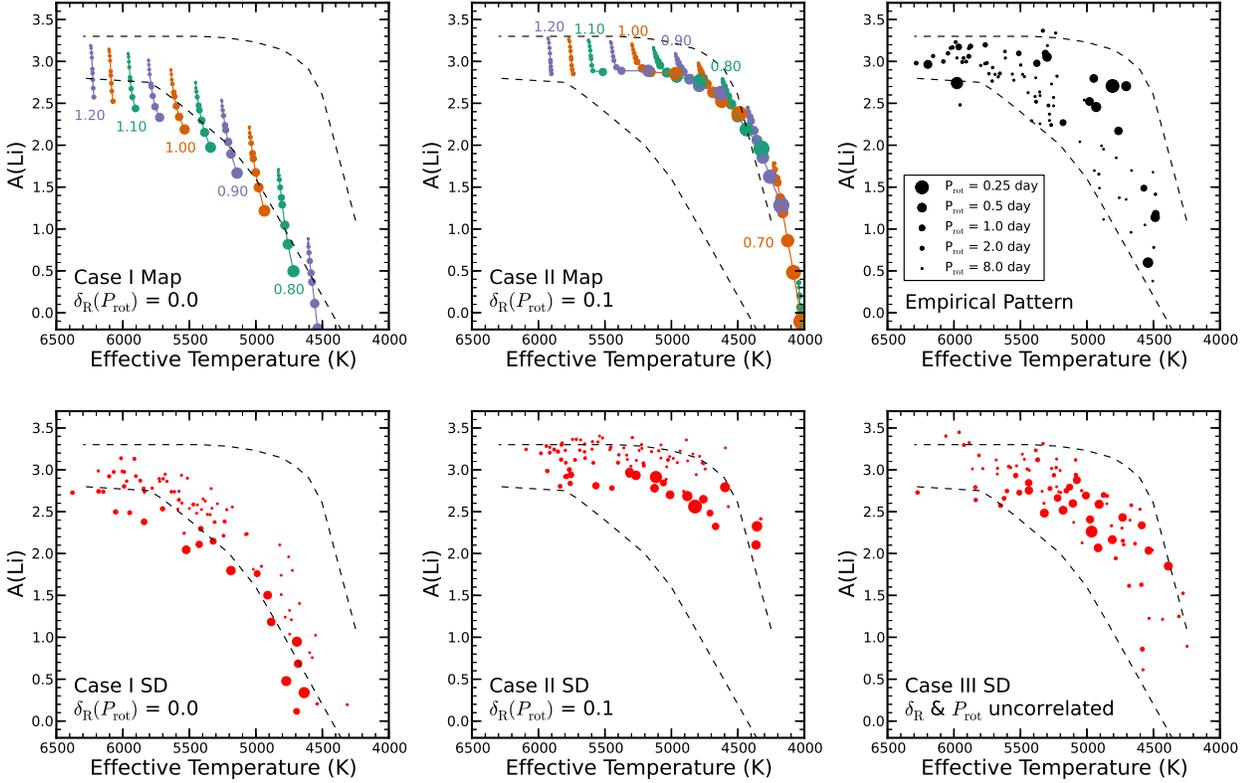}
\caption{Maps and synthetic distributions for the three limiting cases described in $\S$\ref{sec:case1}--\ref{sec:case3}. The lines in the maps are equal mass, and the points represent different disc-locking times, whereas each point in the synthetic distribution represents a forward modeled Pleiad. The size of each point reflects the rotation rate at 125 Myr, as indicated by the key in the upper right panel. Case I assumes that all stars, regardless of rotation, have solar calibrated mixing lengths. Case II assumes that all stars possess 10 per cent radius anomalies, as might be expected from uniform saturation. Case III assumes no correlation between rotation and inflation. Each of these cases fails to reproduce the observed pattern, shown in the upper right panel and represented by approximate envelopes in each of the figures. }
\label{fig:limitingcases}
\end{figure*}

Next, we derive initial conditions by backwards modeling the empirical Pleiades pattern. For each Pleiad shown in Fig. \ref{fig:LiRot}, we determine the $M$ and \td\ which maps onto the current day \teff\ and \prot, given a specific inflation law, and adopt the results as our starting model distribution. We then calculate the Li abundance at 125 Myr for the derived $M$ and \td\ of each Pleiad, given the above rotation/inflation relationship. This method is advantageous: since the synthetic cluster matches the temperature and rotation distribution of the present day Pleiades by construction, a comparison between the Li/rotation pattern of the synthetic cluster and the empirical pattern is unbiased by the sampling of the initial conditions. Finally, we randomly generate \teff\ and A(Li) errors from Gaussians with $\sigma$ = 100K and 0.1 dex, respectively, and apply them to the models to simulate noise.

Using this methodology, we generate two limiting cases which represent distinct classes of physical models: no radius anomaly in any stars, regardless of rotation (I); constant radius anomaly of 10 per cent in all stars, regardless of rotation (II). We further test a third limiting case, with radius anomalies in the range 0--10 per cent, uncorrelated with rotation (III). The methodology of Case III differs somewhat from Cases I and II, since no direct mapping exists between rotation and inflation in the uncorrelated scenario (see $\S$\ref{sec:case3}). With the synthetic distributions in hand, we search for three features of the empirical Pleiades pattern: the median abundance as a function of \teff, the width of the Li distribution as a function of \teff, and a positive correlation between Li abundance and rotation rate. In each case, the resulting distribution is found to be at odds with the empirical Pleiades pattern. Finally, we demonstrate that a power law relationship between \prot\ and \delR\ above a critical rotation period produces a qualitatively accurate synthetic distribution, and show that the agreement is insensitive to the details of the assumed mapping. We caution that the mass dependence, the normalisation, and indeed even the existence, of the correlation between rotation and inflation are at present unresolved (see $\S$\ref{sec:mechanism}). We therefore present this section as an exploratory exercise, and do not attempt a rigorous quantitative analysis.

\subsection{Case I: no inflation} \label{sec:case1}
The first limiting case assumes that no inflationary process operates on the pre-MS, and all stars can be described with a solar calibrated mixing length -- hence \delR(\prot) $\equiv$ 0.0. This represents the fiducial prediction of stellar evolution theory.

The top left panel of Fig. \ref{fig:limitingcases} contains the Case I `map', which shows the values of A(Li), \teff, and \prot\ resulting from each $M$ and \td\ grid point at the age of the Pleiades. In this plot, lines connect models of equal mass but different disc-locking time. The size of the points denote the rotation rate, and the black dashed lines mark the approximate envelope of Pleiades data (see upper right panel). Line colours alternate to ease the distinguishing of mass tracks. Just as in the pure rotation models of $\S$\ref{sec:roteffects}, the stars with the shortest disc-locking times attain the most rapid rotation, and show the greatest degrees of Li depletion. The models bracket the lower envelope of the Pleiades distribution, but do not approach the upper envelope. 

The synthetic distribution for this case, generated as described above, appears in the bottom left panel of this figure. As can be seen, the resulting Li pattern is inconsistent with empirical cluster distribution. The median abundance falls below the empirical median, the distribution width fails to increase substantially towards lower \teff, and while the slowly rotating models approximately correspond to the locus of slow rotators in the Pleiades, the rapid rotators are too strongly depleted in Li to reproduce the observed pattern. As none of the aforementioned empirical features is reproduced by this limiting case, we strongly rule out the possibility that the observed Li pattern results purely from rotationally-induced mixing.

\subsection{Case II: uniform inflation} \label{sec:case2}
The second limiting case assumes a `saturated' inflation law; all stars possess the same radius anomaly regardless of rotation rate -- hence \delR(\prot) $\equiv$ 0.1. This scenario is consistent with \citet{jackson14}, who found that the \textit{average} K dwarf radius in the Pleiades is $\sim$10 per cent larger than expected from isochrones calibrated on old, inactive stars. Furthermore, uniform \delR\ on the pre-MS is supported by \citet{preibisch05} and \citet{argiroffi13}, who found that nearly all stars in the pre-MS associations ONC and h Per are saturated, and might therefore have the same radius anomaly if magnetic field proxies correlate directly with radius inflation.

The upper middle panel of Fig. \ref{fig:limitingcases} shows the mapping for this scenario. The slowest rotating models tend to lie above and to the right of their counterparts in Case I, as inflation suppresses Li destruction and reduces the surface temperature. For the rapidly rotating stars, the decrease in \teff\ is far more extreme, causing a strong shift to the right. The result is a substantial overlap of the mass tracks. This is not seen in the most massive tracks, since even the fastest rotators in these bins have spun down substantially by 125 Myr (see Fig. \ref{fig:InflTracksRot}). The overall trend is a strong suppression of Li destruction in all stars, regardless of rotation.

The lower middle panel shows the resulting synthetic distribution. As in Case I, the models occupy a band at fixed \teff\ that is far narrower than the empirical spread. Unlike Case I, the median abundance trend lies far above the empirical median, and traces the upper envelope of the Pleiades distribution. This signifies that a uniform suppression of depletion is too strong to reproduce the most Li-poor stars below $\sim$ 5500K. The location of the rapid rotating models approximately corresponds with the rapid rotators in the Pleiades, but they remain more depleted than slowly rotating models. Once again, the qualitative features of the Pleiades pattern are not in general reproduced, ruling out uniform inflation from the epoch of pre-MS Li destruction to the present. We note however that Case I and Case II suggestively bracket the observed data range.

\subsection{Case III: uncorrelated inflation} \label{sec:case3}
The final limiting case assumes that no direct correlation exists between rotation and inflation. As stated above, a direct mapping such as those constructed for Cases I and II cannot be defined for Case III, as rotation and inflation are by definition uncorrelated. Instead, we adopt the initial distribution of $M$ and \td\ derived from the Pleiades for Case I, randomly draw an inflation factor between 0--10 per cent for each individual star, and interpolate in the model grid to find the 125 Myr distribution. A physical scenario which may correspond to this limiting case is proto-stellar radius variations caused by accretion ($\S$\ref{sec:mechanism}).

The resulting synthetic distribution is presented in the bottom right of Fig. \ref{fig:limitingcases}. Unlike the previous two cases, the model dispersion now matches the empirical dispersion reasonably well. In particular, it increases from $\sim$0.5 dex at the solar temperature to $\sim$1--1.5 dex below 5000K, in good agreement with the Pleiades. The median of the pattern appears to track the empirical median as well. However, the rotation distribution does not correspond to the observed pattern. Since rotation and inflation are uncorrelated, we see far too many slow rotators at the top of the distribution, and far too many rapid rotators at the bottom. For example, cool-ward of 5000K, 21 out of the 32 stars with rotation periods greater than 2 days have A(Li) $>$ 2 in our synthetic distribution (65 per cent), compared to only 1 out of 15 in the empirical distribution. The probability of finding only 1 slow rotator out of 15 when the expectation is 65 per cent is $\sim 5 \times 10^{-4}$, ruling out Case III at high confidence. This suggests that without some correspondence between rotation and inflation, the Li--rotation correlation in the Pleiades cannot be produced.

\subsection{Case IV: correlated inflation} \label{sec:case4}
None of the first three cases resemble the observed pattern, but they collectively point towards a resolution. As noted, the locus of slow rotators in the \teff--A(Li) plane is accurately predicted by Case I, the locus of rapid rotators is accurately predicted by Case II, and the \teff\ dependence of the median and width are produced by a range of radius anomalies from 0--10 per cent, as in Case III. If the Li depletion history of slow rotators was similar to those in the Case I mapping (i.e. no radius anomaly), and the depletion history of rapid rotators was similar to those in the Case II mapping (i.e. $\sim$10 per cent radius anomaly), the Li--rotation correlation in the Pleiades would be produced, and the \teff\ dependence of the median and dispersion may be preserved. Following this logic, our last class of models assumes a correlation between rotation rate and inflation factor, such that rapid rotators are physically larger than slow rotators.

\begin{figure*}
\includegraphics[width=7.0in]{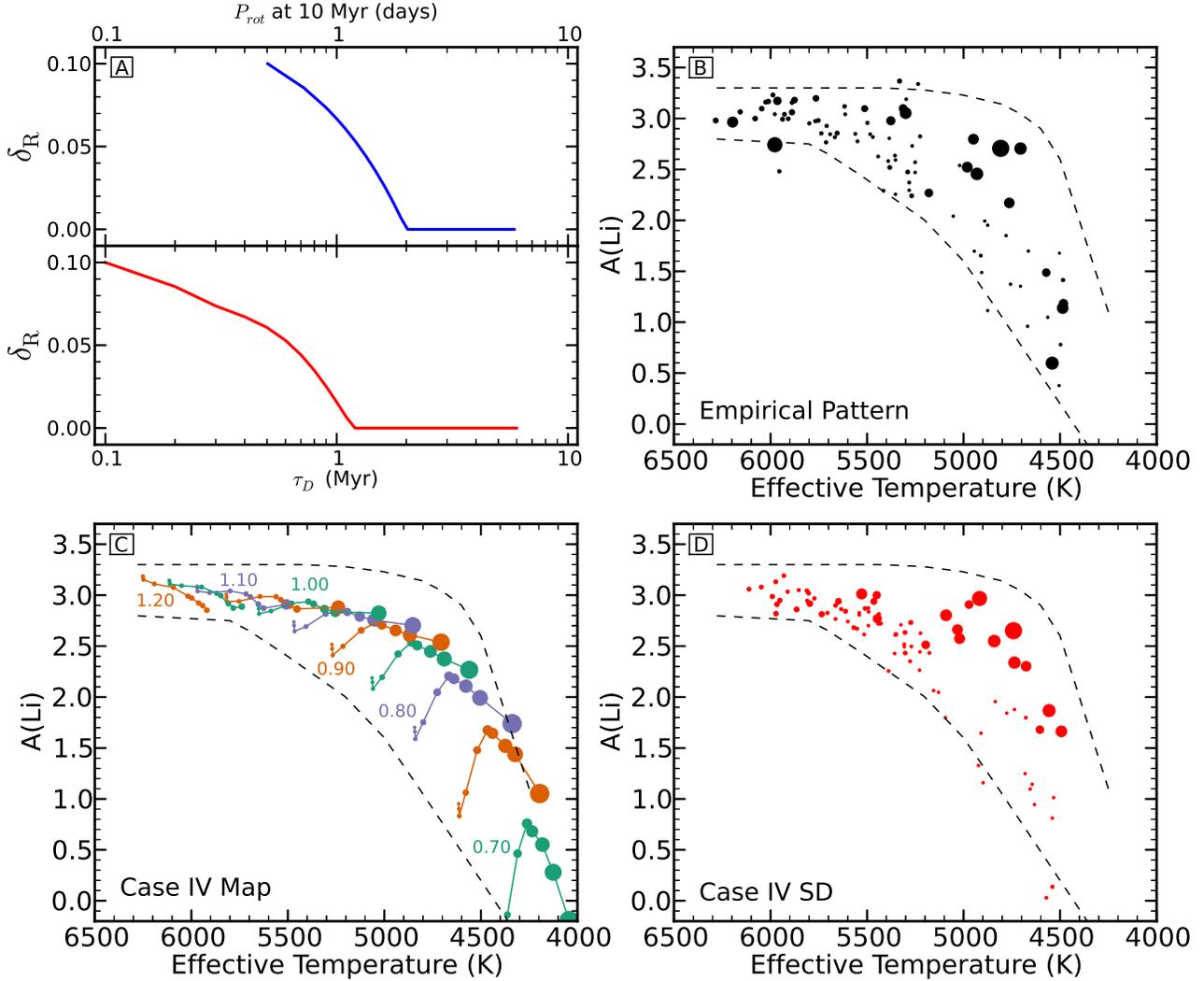}
\caption{The results of assuming a linear scaling between rotation and inflation ($\beta$ = 1, P$_{0\%}$ = 2 days in Eq. \ref{eqn:inflate}). \textit{A:} A graphical representation of the inflation function. The top panel shows the mapping between rotation at 10 Myr and the radius anomaly, and the bottom shows how this maps on to our distribution of disc-locking times. \textit{B:} The empirical Pleiades pattern reproduced. \textit{C:} The map resulting from this inflation paradigm. The effect of more rapid rotation on Li is mass dependent and non-linear. The general trend is to push rapidly rotating stars to the upper right portion of this panel, near the empirical upper envelope shown in dashed black. \textit{D:} The synthetic distribution resulting from this mapping. This paradigm reproduces the empirical median, the empirical distribution as a function of \teff, and the observed Li--rotation correlation.}
\label{fig:linear_map}
\end{figure*}

\begin{figure*}
\includegraphics[width=7.0in]{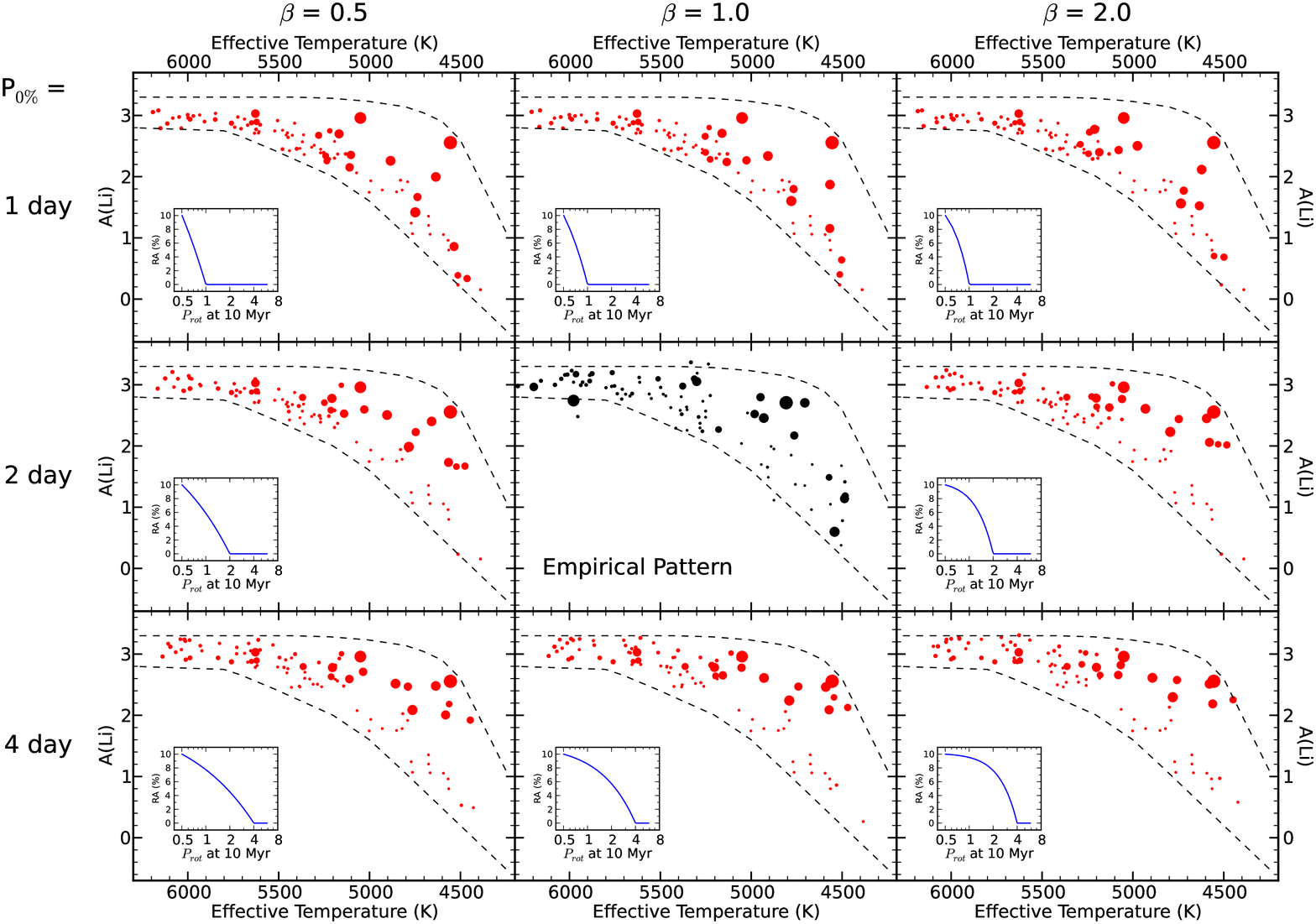}
\caption{Synthetic distributions resulting from a variety of choices of $\beta$ and P$_{0\%}$, along with the individual mappings between rotation and inflation illustrated in inset panels. The qualitative features of the pattern shown in Fig. \ref{fig:linear_map} are a general feature of power law inflation models, but the exact distribution is sensitive to the details of the equation and the underlying rotation distribution. The value of P$_{0\%}$ has a stronger effect on the distribution than $\beta$, since the fraction of stars in the rapidly rotating locus depends on the absolute number of highly inflated members.}
\label{fig:many_clusters}
\end{figure*}

We implement Case IV as follows. The maximum radius anomaly in our grid (10 per cent) is assigned to the stars that achieve \prot\ $\sim$ 0.5 days at 10 Myr, our fastest rotators. The radius anomalies of slower rotators are then determined by a power law, which extends down to a cut-off rotation period, below which stars are un-inflated. The functional form is given by Eq. \ref{eqn:inflate}.

\begin{equation}
\delta_{\rm R} =
\begin{cases}
0.1 \times \left( P_{0\%}^\beta - P^\beta \right) / \left(P_{0\%}^\beta - P_{10\%}^\beta \right), & P \leq P_{0\%} \\
0.0, & P > P_{0\%}
\end{cases}
\label{eqn:inflate}
\end{equation}

Here, $P$ is the model rotation period at 10 Myr, $P_{10\%}$ is the rotation period at which stars attain the maximum radius anomaly of 10\% (set to 0.5 days as described above), and $P_{0\%}$ is the rotation period below which radius inflation is absent. $\beta$ represents the power-law exponent of the relation. Thus, stars rotating slower than $P_{0\%}$ at 10 Myr will have no radius inflation, stars rotating faster than $P_{0\%}$ but slower than $P_{10\%}$ at 10 Myr will have an inflation factor between 0 and 10 per cent, and stars spinning at $P_{10\%}$ at 10 Myr will have a 10 per cent radius inflation. An example of the inflation law is shown in blue in panel A of Fig. \ref{fig:linear_map} (see below for details).

While this formulation is clearly \textit{ad hoc}, its features are motivated by data. Rapidly rotating, magnetically active stars tend to show the greatest deviations from standard models \citep[e.g.][]{lopez-morales07}, whereas slowly rotating, inactive stars generally agree with stellar models. This suggests that measurable inflation sets in above a critical rotation rate, as we have modeled. A power law form was motivated by the power law dependence of chromospheric activity on rotation \citep[e.g.][]{pizzolato03}, and our maximum anomaly was chosen to coincide with the upper envelope of claimed anomalies in the literature, as previously stated. 

With this methodology, we first test a linear scaling ($\beta$ = 1), with a transition period $P_{0\%}$ = 2 days. This version of Eq. \ref{eqn:inflate} is shown in blue in panel A of Fig. \ref{fig:linear_map}, and the corresponding mapping between \td\ and \delR\ is shown in red. Panel C shows the map for Case IV. The behaviour of the models is striking. Unlike Cases I and II, where a shorter \td\ always results in a lower abundance, the response of Case IV models to shorter \tds\ is mass dependent. Higher mass models (e.g. 1.2\msun) always cool down and deplete more with greater rotation, whereas lower mass models (e.g. 0.8\msun) show a more complicated relation. For long \td\, greater rotation induces more depletion, since none of these models are inflated in our chosen mapping. For intermediate \td\, greater rotation actually \textit{inhibits} depletion, since intermediate rotation (\td\ $\sim$ 0.7 Myr) has a smaller effect on Li depletion than intermediate inflation (\delR\ $\sim$ 0.05; see Fig. \ref{fig:InflTracksInf}). Finally for short \td, mixing again dominates, and the Li abundance decreases with faster rotation. The overall trend is that shorter disc-locking times drive the models to the upper right of this figure, and increase their Pleiades-age rotation rate.

The Case IV synthetic distribution derived from the empirical Pleiades pattern appears in panel D. Unlike the first three cases, several salient qualitative features of the empirical distribution are reproduced by this mapping. First, the locus of slow rotators roughly corresponds to the lower envelope of the Pleiades distribution, just as in Case I. Second, the most rapid rotators lie above and to the cool side of this locus, roughly corresponding to the locus of rapid rotators in the empirical pattern, as in Case II. This morphology is due in part to the decreased rate of Li destruction on the pre-MS, and in part to the reduced \teff\ of rapidly rotating, inflated stars. Third, a significant dispersion sets in around 5500K and increases towards cooler temperatures, in excellent agreement with the Pleiades. It can be understood why this occurs at 5500K from the Case IV map. Rapid rotation displaces stars horizontally towards cooler temperatures in the \teff--A(Li) plane; this horizontal path coincides with the slow rotators above 5500K, but lies above the slow rotators at lower temperatures, since these stars are strongly depleted. The \teff\ decrement due to rotation and inflation is of a similar magnitude for all masses, but the effect on the observed distribution is more apparent in the sub-solar regime. One location where the models conspicuously fail is the coolest rapid rotators; four stars near 4500K are predicted to have A(Li)~$>$~1.5 by our models, but are actually more heavily depleted by 0.5-1 dex. This may reflect a deficiency in our modeling of radius anomalies, which does not account for changes as a function of mass and evolutionary state. For instance, the true maximal radius anomaly could decrease with mass, so that depletion factors among the fastest spinning K dwarfs are larger than we have predicted.

Despite the \textit{ad hoc} nature of our \delR(\prot) scaling, this model succeeds at accurately predicting several puzzling features of the observed Pleiades distribution. However, one could reasonably contend that fine tuning may be required to achieve this qualitative agreement. To address this concern, we assess the generality of the Case IV agreement by calculating synthetic cluster distributions resulting from different choices of $\beta$ and $P_{0\%}$ in Eq. \ref{eqn:inflate} (Fig. \ref{fig:many_clusters}). From left to right we increase the power law exponent ($\beta$), and from top to bottom we decrease the rotation rate at which inflation sets in ($P_{0\%}$). The initial conditions were derived for each individual case using the specified inflation prescription, and the same run of \teff\ and A(Li) errors was applied to each. From this plot, we see that some features of the Case IV mapping are generic, such as the population of rapid rotators to the upper right of the main locus of slow rotators, and the onset of dispersion at $\sim 5500$K. Other features of the synthetic pattern vary from mapping to mapping, such as the number of rapid rotators that lie near the bottom of the abundance distribution below 5500K (approximately half for $P_{0\%} = 1$ day, none for $P_{0\%} = 4$ days; the dependence on $\beta$ is negligible). Never the less, since the important qualitative features of the Pleiades pattern are generically reproduced by each considered version of Eq. \ref{eqn:inflate}, we conclude that fine tuning is not essential for the conclusions of this section. 

To summarise $\S$\ref{sec:case4}, the three qualitative features of the empirical pattern described at the beginning of $\S$\ref{sec:section4} are reproduced by a relationship between rapid rotation and inflation, and these qualitative features are largely insensitive to the exact choice of mapping function. We conclude that Case IV strongly outperforms each of the limiting cases, and produces an observational pattern consistent with the Pleiades.

\section{Discussion} \label{sec:discussion}

We have presented the first consistent physical model that replicates the qualitative pattern of Li and rotation observed in the Pleiades. $\S$\ref{sec:case4} delineated the properties of a successful solution, so we now turn to plausible explanations for the rotation--inflation correlation that we have inferred. We consider both the inhibition of convection by magnetic fields, and the impact of mass accretion during the proto-stellar phase. Next, we discuss a method for detecting radius dispersions on the pre-MS and beyond, as a test of our scenario and as a means to constrain the putative connection between rotation and inflation.

\subsection{The mechanism of inflation} \label{sec:mechanism}

Our paradigm assumes the existence of a mechanism which reduces the temperature at the base of the convection zone across the observed range of rapid rotators on the pre-MS. This behaviour naturally results from an increase in the stellar radius, so we have developed our models in the context of radius anomalies correlating with rotation. There is observational support for this effect from young clusters and associations \citep{littlefair11,cottaar14}, and from pre-MS eclipsing binaries \citep{stassun07,stassun14}. In this section, we discuss two possible mechanisms for inducing a rotation-radius correlation.

We have modelled radius inflation as a product of inhibited convective efficiency. This may result as a direct consequence of rapid rotation, when extreme Coriolis forces begin to govern the behaviour of convective particles \citep[e.g.][]{meisch00}, but we focus on the suppression of convection produced by rotationally-generated magnetic fields. Magnetic fields in the envelopes of low mass stars can disrupt the motions of charged particles when they attempt to cross field lines. This leads to an increase in the convective energy flux per unit area in the non-magnetic regions, forcing the radius to expand in order to match the flux emanating from the core \citep{spruit82,andronov04,chabrier07,feiden13,feiden14}. 
An important consequence of this effect is the formation of star spots on the surface, which result from the inefficient heat flux induced by magnetism. Several authors have taken this approach, and considered the impact of spots on the radii of stars \citep[][Somers \& Pinsonneault 2015, in prep]{chabrier07,jackson09,jackson13,jackson14}. Polytropic models show that a filling factor of $\sim$0.35--0.51 is sufficient to achieve a K dwarf inflation $\sim$10 per cent, consistent with filling factors $\sim$0.2--0.4 found on active G and K dwarfs \citep{o'neal04}. This behaviour appears particularly effective in stars with deep convection zones \citep{jackson09}, and so may be effective on the pre-MS.
Magnetic fields can also provide pressure support in stellar interiors, thus working to counteract the force of gravity similar to rotational support \citep{mullan01,macdonald09,macdonald12,macdonald13}. Each of these effects serves to puff the radius up in stellar models, leading to a reduced temperature in the convection zone.

While these effects would naturally produce the type of correlation we require, activity studies of pre-MS clusters have found that most young stars of ages $\sim$ 1--15 Myr are saturated, meaning that the strength of their magnetic field proxies (i.e. X-rays) no longer correlate with rotation rate \citep{preibisch05,argiroffi13}. The predominate interpretation of this result is that magnetic fields cease to increase in strength beyond the critical Rossby number at which saturation sets in \citep[$R_N = P_{\rm rot}/\tau_{CZ}\sim 0.1$;][]{pizzolato03}. If true, it would appear difficult for convection to be progressively inhibited by rapid rotation among pre-MS stars. However, an alternative interpretation is that magnetic field proxies decouple from the strength of the surface field at the onset of saturation, so that magnetic fields can continue to grow with rotation without impacting proxies. For instance, saturation could represent a threshold above which chromospheric and coronal heating cannot increase, but the spot filling factor can \citep[e.g.][]{odell95}. Alternatively, the spot filling factor could saturate, but the spot temperature could continue to decrease with greater rotation. Several studies have attempted to measure directly the magnetic field strengths of saturated stars and have found mixed results \citep[e.g.][]{saar96,saar01,reiners09}, but their efforts are hampered by the extreme difficulty of inferring Zeeman broadening from stars whose lines are already highly broadened by rotation \citep{reiners12}. Due to these observational issues, the continued inhibition of convection with greater rotation in the saturated regime remains unresolved at present, but further work in this area will ultimately favour or disfavour this mechanism as a viable avenue.

Mass accretion may also play an important role in determining pre-MS radii. Proto-stars accrete the majority of their mass during the first $\sim10^5$~yr, when the initial core of $M \sim 10^{-3}$\msun\ gains material from the surrounding over-density in its natal molecular cloud. Accretion likely progresses at a modest level ($10^{-8}$--$10^{-7}\ {\rm M}_{\odot}$ yr$^{-1}$), interspersed with short bursts of vigorous accretion ($10^{-6}$--$10^{-4}\ {\rm M}_{\odot}$ yr$^{-1}$) until the final mass is assembled \citep[see][]{baraffe12}. A major unknown factor in this picture is the amount of thermal energy deposited during these intense accretion bursts. `Cold accretion' scenarios assert that all of the accreted thermal energy is radiated away at the photosphere, so only mass (and not entropy) is added to the proto-star \citep[][and references therein]{hartmann97}. This additional mass leads to a contraction of the radius. Alternatively, `hot accretion' scenarios suggest that a substantial fraction of the thermal energy of the accreted material is absorbed into the stellar envelope. This type of accretion can increase stellar radii \citep[e.g.][]{palla92}.

In either case, a dispersion in radii will arise from differing accretion histories, and may persist throughout the pre-MS due to the lengthy thermal readjustment times. This leads to a range of Li depletion efficiencies during the epoch of standard burning \citep{baraffe10}. If hot accretion is accompanied by substantial angular momentum deposition, or if the rotation period of cold accretors remains fixed during radius contraction due to magnetic interactions with circumstellar material, then the necessary rotation correlation will be likewise present. This scenario makes two interesting predictions. One, that the rotational history of stars is intrinsically linked to their accretion history. Two, that radius inflation is only a transient event on the pre-MS, and therefore will be gone by the age of the Pleiades, leaving behind a Li dispersion as a fossil record of early time inflation. The debate in the literature about hot versus cold accretion scenarios remains unsettled \citep{baraffe09,hosokawa11,baraffe12}, so further work remains to determine precisely how accretion impacts stellar radii on the early pre-MS.

While these mechanisms are fundamentally different, their impact on the structure at the base of the convection zone will be analogous. Any increase in the stellar radius moves the base of the convection zone outwards, reducing the rate of Li depletion, and any decrease in the stellar radius has the opposite effect. We therefore believe that the results of $\S$\ref{sec:section3}--\ref{sec:section4} hold in the presence of a radius dispersion, regardless of the underlying cause. Given recent efforts to probe the radius--rotation connection in young stars, constraints may soon be placed on the connection between these two properties during the pre-MS. 

\subsection{Testing the inflation paradigm} \label{sec:testing}

\citet{jackson14} claimed that the average radius of the Pleiades K dwarfs is inflated by $\sim$10 per cent, but they did not attempt to characterise the distribution about this average. If a radius dispersion was present in the Pleiades at $\sim$10 Myr, as required by our theory, then a remnant of the dispersion might still exist today. Detection of a radius dispersion at 125 Myr would have important implications not only for the evolution of Li, but for the structure of stars on the pre-MS, the behaviour of saturated stars, and the effects of spots and magnetic fields on stellar radii.

A viable avenue for testing whether a spread is present in the Pleiades is through obtaining surface gravities for a large number of cluster members. If rapidly rotating Pleiads are physically larger than slowly rotating Pleiads, a dispersion in \logg\ at fixed colour arises due to three compounding effects. First, a radius increase of $\sim$10 per cent naturally reduces the surface gravity (g $\propto$ M/R$^2$) compared to non-inflated stars at fixed colour. Second, the most inflated stars are very rapidly rotating; rapid rotation reduces \teff\ at fixed mass (Fig. \ref{fig:InflTracksRot}), enhancing the effect at fixed temperature. Third, since inflation also reduces \teff, the most inflated stars at fixed colour are also the most massive. Surface gravity decreases with increasing mass, because R is proportional to M in the solar regime (g $\propto$ M/R$^2$ $\propto$ M/M$^2$ $\propto$ 1/M), leading to an additional reduction in the $\log g$ of the most inflated (most massive) stars at fixed colour. 

For a 10 per cent dispersion in radius, our models predict that the \logg\ dispersion could be as large as 0.2 dex at fixed colour. This signal may be detectable through granulation-induced light curve flicker \citep{bastien13}. The characteristic time-scale of the formation and dissolution of convective cells on the surface of stars is proportional to the surface gravity. This induces a correlation between the power of short time-scale luminosity variations and the strength of surface gravity, permitting the inference of stellar \loggs\ with an accuracy of 0.1--0.15 dex \citep[e.g.][]{bastien13,bastien14}. This signal can be extracted from high precision, long baseline, short cadence photometry, such as that obtained by the $Kepler$ satellite. Although its mission has ended, $Kepler$'s descendant $K2$ will observe a field containing the Pleiades in early 2015 \citep{howell14}, and despite the degraded photometric quality, may be able to detect granulation flicker at the necessary level. This offers a unique opportunity to detect and characterise the distribution of surface gravities, and search for a dispersion about the inflated mean. 

This test will open the door for using Li abundances to understand differential pre-MS stellar structure, but much work remains to be done in order to correctly interpret the information. First, we must determine the mechanism driving the putative connection between rotation and inflation. If magnetic fields are responsible, how do they impact stellar structure in the saturated regime? If accretion is responsible, is the cold or hot scenario more appropriate, and how do they impart the angular momentum distribution? If differential radius inflation on the pre-MS is not responsible for the observed Li dispersion, then how is it formed? Second, we must understand the mass, rotation, and age dependence of radius inflation on the MS. $K2$ will place constraints on this mechanism, since it will measure surface gravities in a large number of clusters of different ages, and track the evolution of anomalies along the MS. Third, the systematic effects plaguing the precise determination of Li abundances and effective temperatures in rapidly rotating K dwarfs must be resolved. These include the shifting spectral energy distributions driven by surface spots, the blending of the Li absorption line with nearby lines due to rotation, and the effects of active chromospheres on line formation \citep[see][and references therein]{king10}. Once accurate \teffs\ and abundances can be determined, the true rate of Li depletion for individual stars on the pre-MS can be derived, and the magnitude of the Li dispersion at fixed mass, and hence the degree of radius inflation on the pre-MS, will be revealed. Finally, observations of the surface gravity distribution of the Pleiades can only provide circumstantial evidence in favour of our theory. Detecting a dispersion in radii correlated with rotation in a $\sim$10 Myr old cluster, or detecting a correlation between rotation at Li abundance at fixed temperature in a 10--20 Myr cluster, will serve as the decisive test of an inflationary origin for the observed Li--rotation correlation. 

\section{Summary and conclusions} \label{sec:conclusions}

In this paper, we have explored the simultaneous impact of rotation and radius inflation on the depletion of lithium from stellar models during the pre-MS. Our goal is to describe the generic features of a mechanism which would bring the predictions of rotational mixing into concordance with the well-known Li--rotation correlation in the K dwarfs of the Pleiades. Such a mechanism must suppress Li destruction in rapid rotators relative to slow rotators, thus overcoming the trend predicted by rotational mixing calculations. As radius anomalies $\sim$10 per cent are sometimes found in rapidly rotating stars, and a increased radius is known to inhibit Li depletion in standard models, we are motivated to test whether a correlation between rotation and radius inflation on the pre-MS (induced by some physical process neglected in standard stellar models) could be responsible for the observed pattern.

We first reviewed the impact of rotation and inflation on the evolution of Li abundances. In accordance with previous studies, we find that the depletion of Li is a strong function of both properties, such that rotation \textit{increases} the rate of Li destruction, and radius inflation, modeled by a reduction of the mixing length parameter, \textit{decreases} the rate of Li destruction. While this behaviour is straight forward, the picture becomes far more complicated when the two effects are combined. Rapid rotation and substantial radius inflation (10 per cent) enhances Li depletion for stars with $M > {\rm M}_{\odot}$, but inhibits Li depletion for stars with $M < {\rm M}_{\odot}$, relative to non-rotating benchmarks. This is the first new result of our work, and implies that if the radii of just the rapid rotators in the Pleiades were inflated by 10 per cent during the pre-MS, while the radii of the slow rotators agreed with standard predictions, then a positive correlation between the rate of rotation and the abundance of Li at 125 Myr could arise in K dwarfs, in agreement with observations.

Next, we assessed whether this qualitative notion could explain the observed relationship between Li and rotation in the Pleiades. To do this, we generated synthetic cluster distributions derived from the empirical \teff/rotation distribution of the Pleiades which obey various inflation/rotation prescriptions, and compared them to the empirical Pleiades pattern. First, we found that three limiting cases, which represent distinct classes of physical models, were unable to reproduce the Pleiades pattern: no radius inflation in all stars regardless of rotation, uniform radius inflation of 10 per cent in all stars regardless of rotation, and radius inflation uncorrelated with rotation. Next, we adopted a prescription where rapidly rotating stars are more inflated than slowly rotating stars, and found that observed pattern in the Pleiades is generically produced. The dispersion in Li remains small around the solar temperature and increases to greater than an order-of-magnitude towards cooler temperatures, precisely as is found in the Pleiades. Moreover, the median abundance trend is accurately reproduced, and rapidly rotating stars are found to be more rich in Li than slowly rotating stars in the K dwarf regime. Furthermore, we found that these qualitative features are insensitive to the precise details of the rotation/inflation prescription we adopt as long as more rapid rotation leads to greater inflation, demonstrating that fine tuning is not required to achieve this level of agreement. This is the first consistent physical explanation for the Pleiades Li pattern to date, so we conclude that it is a strong candidate for explaining the observed pattern.

After establishing that a connection between rotation and inflation could explain the Pleiades pattern, we discussed what sort of physical mechanisms might lead to this correlation. The inhibition of convection due to strong magnetic fields in rapid rotators could lead to such a correlation if field strengths continue to grow with rotation in the saturated regime. Accretion could also alter the radii and rotational histories of proto-stars, plausibly inducing the observed correlation. Finally, we suggest that the remnants of a pre-MS radius dispersion may still be present in the Pleiades, and a survey of stellar surfaces gravities in this cluster could place strong constraints on the radius dispersion needed at 10 Myr to produce the observed pattern. The ongoing $K2$ mission offers a valuable opportunity to perform this experiment.

We would like to thank the anonymous referee, whose suggestions improved this paper significantly. This work was supported through the NSF grant AST-1411685.

\end{document}